# HUMAN EMOTION RECOGNITION FROM AUDIO AND VIDEO SIGNALS

Sai Nikhil Chennoor[1], B. R. K. Madhur[2], Moujiz Ali[3], Dr. T. Kishore Kumar[4]

[1,2,3,4] Department of Electronics and Communication Engineering,

National Institute of Technology, Warangal, Telangana, 506004, India

*Abstract* — The primary objective is to teach a machine about human emotions, which has become an essential requirement in the field of social intelligence, also expedites the progress of human-machine interactions. The ability of a machine to understand human emotion and act accordingly has been a choice of great interest in today's world. The future generations of computers thus must be able to interact with a human being just like another. For example, people who have Autism often find it difficult to talk to someone about their state of mind. This model explicitly targets the userbase who are troubled and fail to express it. Also, this model's speech processing techniques provide an estimate of the emotion in the case of poor video quality and vice-versa.

## I. INTRODUCTION

Human emotion is the best to understand the attitude of an individual towards another or a particular situation. In general, Human emotions are a composition of speech and facial expressions. It is a new domain of research that provides the machines with the ability to understand and sense human emotions. Machines must be capable of understanding and compassionately responding to these emotions.

The human face is a complex interaction of various activities, displaying human emotions expressed by several signals. These signals are the variations and orientation of the human facial structure. The facial expressions can describe one neutral and six fundamental emotions (happy, sad, fear, disgust, surprise, and angry). In this paper, we demonstrate a model developed for facial expression recognition, 84% accurate on the JAFEE database [2].

Similarly, speech is a complex signal, so from the speech, we classified a signal emotion by taking some features like pitch, entropy, and energy. These speech features are language independent and non-verbal. The model resulted in an accuracy of 78.94% for speech emotion recognition on the Berlin Emo database [3].

The audio extractor automatically filters out the audio signal from the captured video and audio signals of a single person. Now the video is divided into frames. The emotion of each frame is recognized. The emotion that appears in most of the frames now becomes the video signal's emotion. Similarly, the audio signal's emotion is classified. At the combining stage, we make a decision, and overall emotion from audio and video signals is recognized. The block diagram of the system appears in Fig .1.[1][4].

## II. FACIAL EXPRESSION RECOGNITION

The best display of Human emotion is through facial expressions. Facial expression recognition methods aim to automatically build a system for the classification of facial expressions from the images. This system will be able to classify one neutral emotion, six fundamental emotions, namely Happy, Anger, Fear, Disgust, Sad, and Surprise. The facial expression recognition model has three stages.
- Face detection
- Feature extraction
- Emotion recognition from the image

The block diagram of the facial expression recognition model appears in Fig.2.

### A. Face Detection

Face detection is to detect the face in the given input image. To search for faces within an image, we use OpenCV and machine learning algorithms. Viola and Jones adapted the idea of using Haar wavelets for face detection and developed the so-called Haar-like features [5]. This algorithm can be able to detect the face with 95% accuracy. In general, for face detection in a given image, we have to consider nearly 6000 features. The OpenCV cascade breaks the problem of detecting faces into multiple stages. So, if the first stage passes, then we go for the second stage and so on. The advantage of this method is that most of the images will not be able to pass all stages, which means the algorithm will not waste time testing all 6,000 features on it. With this algorithm, we can be able to detect the face in an image in real-time [5].

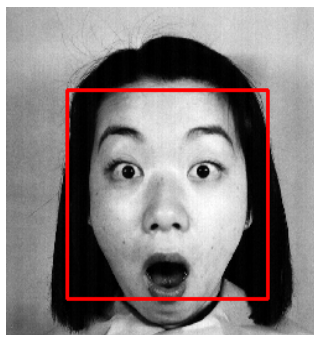

Fig. 3. Detection of the face from the general image

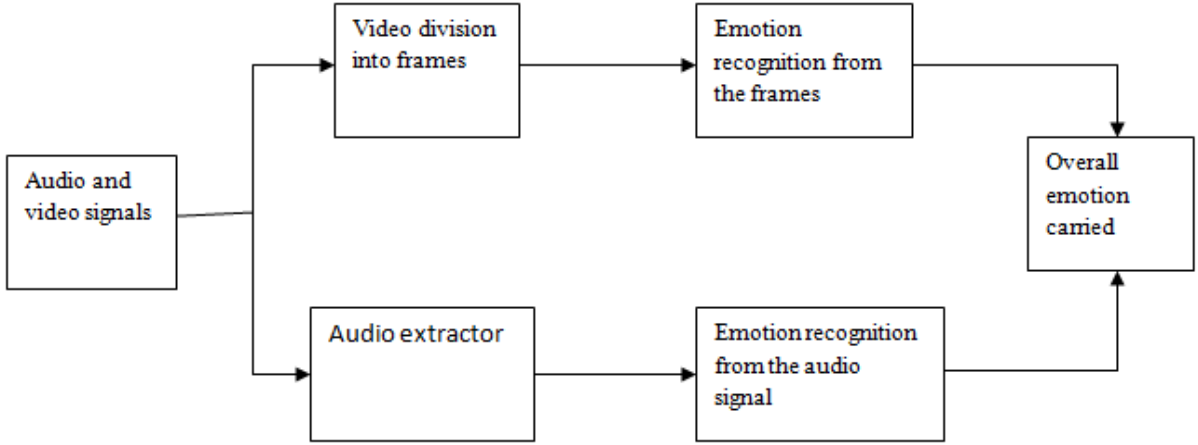

Fig. 1. Block diagram of emotion recognition from audio and video signals.

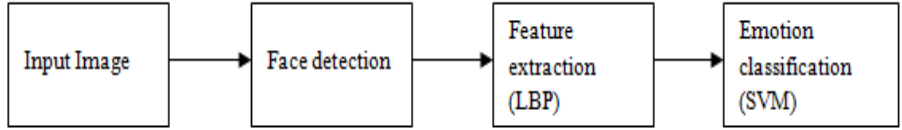

Fig. 2. Block diagram of emotion classification from the facial image.

*B. Feature Extraction*

For feature extraction, we use a local binary pattern. For each pixel in an image, compare the pixel to each of its eight neighbors. We follow the pixels in a clockwise or counterclockwise fashion. Where the center pixel's value is higher than the neighbor's value, write "0". Otherwise, write "1". It gives an 8-digit binary number (which we convert to decimal for convenience). Compute the histogram by taking the frequency of each "number" occurring. Generally, this histogram has 256 bins. However, we now reduce them to 59 bins by using uniform LBP. The conversion from the original image cell to the LBP cell appears in Fig. 4. The binary sequence obtained in the above example is 10011110 when converted to decimal; the value is 158. So, the center pixel of the original image cell is replaced by the decimal value 158. This process repeats for every cell of the original image. The LBP image obtained by the above process appears in Fig. 5.[2].

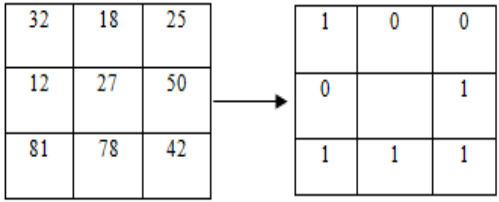

Fig. 4. Conversion from original image cell to lbp cell.

A local binary pattern is a uniform if it contains at most two 0-1 or 1-0 transitions. For example, 00000000(zero transitions) is a uniform pattern, 01010100(six transitions) is not. In the evaluation of the LBP histogram, the histogram constitutes a separate bin for every uniform pattern, a single bin for other non-uniform patterns. The length of the feature vector can be reduced from 256 to 59 using these uniform patterns.

Each image is divided into 7*8=56 sub-regions, as shown in Fig. 6. For each LBP, a histogram is calculated with 59 bins. These calculated histograms are juxtaposed to form a feature vector. So totally, a feature vector of size 3304 is taken and used for emotion classification.

*C. Emotion Classification*

The JAFFE (The Japanese Female Facial Expression Database) database contains 213 images of 7 facial expressions (6 basic facial expressions + 1 neutral) posed by 10 Japanese female models. The facial expression classification model achieved an average accuracy of 84% on this database by using the SVM model. The KNN model resulted in an average accuracy of 75% on the JAFEE database. All these results are tabulated in Table I.

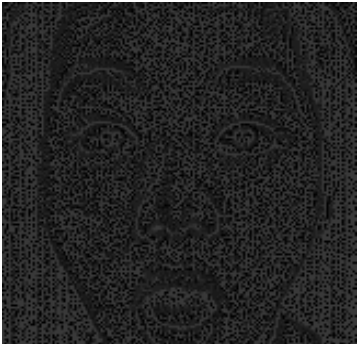

Fig. 5. Image obtained by doing lbp at each cell of original image.

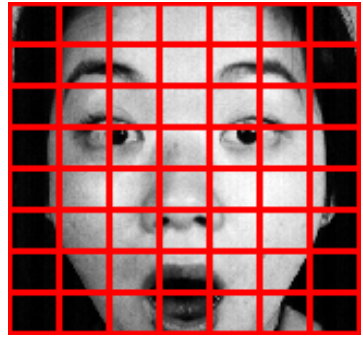

Fig. 6. Image division into 56 regions.

TABLE I. RESULTS OBTAINED FOR FACIAL EMOTION RECOGNITION MODEL ON JAFEE DATABASE.

|     | Training images | Testing images | Training accuracy | Testing accuracy |
| --- | --- | --- | --- | --- |
| KNN | 170 | 42 | 90% | 67% |
| SVM | 170 | 42 | 98% | 75% |
| KNN | 187 | 25 | 90% | 74% |
| SVM | 187 | 25 | 98% | 84% |

## III. SPEECH EMOTION RECOGNITION

One of the most reliable ways of recognizing the emotional state of a person is through facial expressions. Speech, a complex signal that contains information about the speaker, message, and emotions, is another element that can describe human emotion. Speech emotion recognition is a system that recognizes the emotional state of a human being from his or her voice.

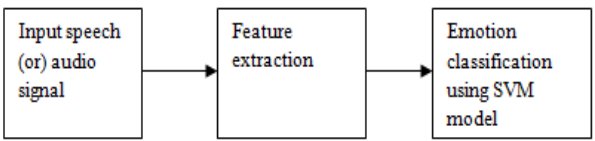

Fig. 7. Block diagram of speech emotion recognition.

### A. Feature Extraction

The sampling frequency of all the audio files is 16KHZ. For feature extraction, we divide the first signal into many frames by maintaining each frame duration of 30ms(approx.), with a 50% overlap. For each frame, we extract 21 features and juxtapose them to form a feature vector. We train the SVM model with this feature vector. This speech emotion recognition model resulted in an average accuracy of 78.94%, upon testing with the Berlin Emo database. We calculate pitch for the entire signal and the remaining features for a frame. Some of these features are in the time domain, and some others are in the frequency domain.

*1) Pitch:* The fundamental frequency of a speech signal is pitch. The calculation of pitch is a complicated task. We use the YIN Pitch Estimation algorithm [8] to detect the pitch values. Let us consider a signal $y_t$ as a periodic function with period T, then by definition.

$$y_t - y_{t+T} = 0 \quad (1)$$

Similarly, for a window of size W, the below expression is valid if $y_t$ is a periodic signal.

$$\sum_{j=t+1}^{j=t+W} (y_j - y_{j+T})^2 = 0 \quad (2)$$

If the signal is not periodic, then for different values of T, we get a difference function as given in eq. 3. Now the value of T where the difference function becomes zero can be used to calculate the pitch of the signal.

$$d_t(T) = \sum_{j=1}^{W}(y_j - y_{j+T})^2 \qquad (3)$$

The above function is zero at T=0. At the remaining values of T, the function is not zero. So a threshold value is set to choose the value of T., But if the limit is set, because of resonance, there can be a chance of producing a series of secondary dips, which might be more profound than the periodic dip. The remedy is to replace the difference function by the "normalized difference function." It is obtained by dividing each value of the difference function by its average over the values. Now a threshold value is set as 0.7. The first dip where the normalized difference function crosses the threshold value can become T. Now sampling frequency divided by the T gives the pitch of the signal.

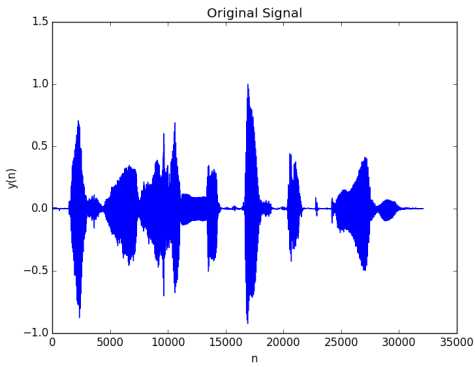

Fig. 8. Speech signal corresponding to happy emotion from Berlin Emo database.

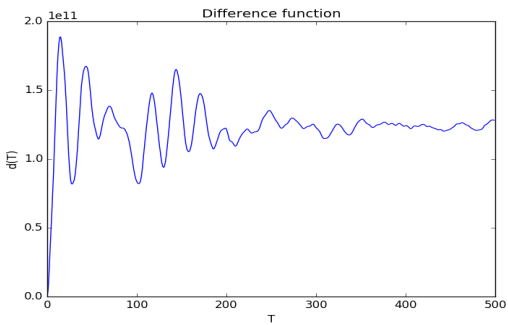

Fig. 9. Difference function.

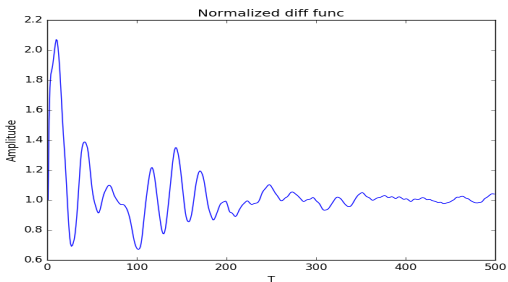

Fig. 10. Normalized difference function.

*2) ZCR:* The zero-crossing rate is the rate at which the signal crosses zero. For a given frame, it is the number of times the signal crosses zero divided by the frame lengths.

*3) Energy:* The energy of a frame is the ratio of the sum of squares of the signal with the length of the frame. The energy of a discrete-time signal y(n), expressed as

$$E_S = \frac{1}{T}\sum_{n=0}^{n=T}(y(n))^2 \qquad (4)$$

Where T is the frame length.

*4) Entropy:* Entropy is a measure of abrupt changes in the signal. Each frame of the signal is divided into subframes. Now let us consider g as the ratio of subframe energy to the total frame energy.

$$g = \frac{E_s}{E_t} \qquad (5)$$

The entropy of energy of the total frame is defined as

$$Entropy = -\sum_{i=1}^{n} g_i * \log(g_i) \qquad (6)$$

Where n is total number of subframes, and $g_i$ is the normalized energy of the sub frame.

*5) Spectral Centroid and Spread:* Spectral centroid indicates the center of the spectrum. It is the ratio of weighted mean of the frequencies in the spectrum, with their magnitudes.

$$centroid = \frac{\sum_{n=0}^{N-1} f(n) * Y(n)}{\sum_{n=0}^{N-1} Y(n)} \qquad (7)$$

Where *Y(n)* represents the weighted frequency value, or magnitude, and *f(n)* represents the center frequency of the bin.

The second central momentum of the spectrum is spectral spread. Let the spectral spread be denoted as ss. Now it can be mathematically defined as

$$ss = \frac{\sum_{n=0}^{N-1}(f(n) - centroid)^2 * Y(n)}{\sum_{n=0}^{N-1} Y(n)} \qquad (8)$$

*6) Spectral Flux:* It basically compares two successive frames. It is the square of the difference between normalized magnitudes of two adjacent frames.

Mathematically spectral flux can be defined as

$$spectral\ flux = \sum_{i=0}^{n}\left(\frac{Y_i}{sumy} - \frac{Yprev_i}{sumprev}\right)^2 \quad (9)$$

Where n is the length of the frame, sumy is the present frame spectral magnitudes sum, sumprev is the previous frame spectral magnitudes sum.

*7) Spectral RollOff:* It is basically the frequency below which 90% of the spectrum is distributed.

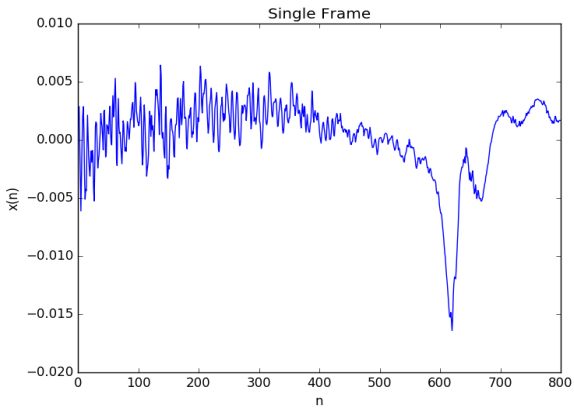

Fig. 11. A single frame of the speech signal.

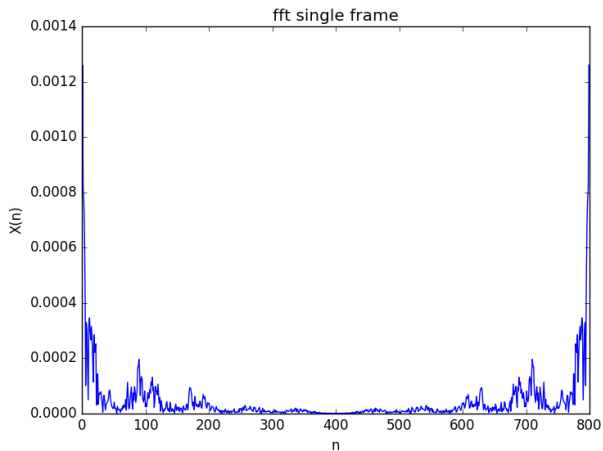

Fig. 12. FFT of the above frame with spectral rolloff=0.997

*8) Entropy of the spectrum:* It is similar to the time domain entropy. Hereafter converting into the frequency domain, each frame is divided into subframes. The term g represents the ratio of the energy of the subframe spectrum to the ratio of the energy of the total frame spectrum.

$$g = \frac{E_s}{E_t} \quad (10)$$

The entropy of the spectrum is mathematically defined as

$$\text{Entropy} = -\sum_{i=1}^{n} g_i * \log(g_i) \quad (11)$$

Where n is the total number of subframes, and $g_i$ is the normalized spectral energy of the subframe.

*9) MFCC:* Mel Frequency Cepstral Coefficients are the most widely used feature in speech emotion recognition. The primary purpose of the MFCC [10] processor is to mimic human ears. The main steps of MFCC, along with the block diagram, are shown in Fig. 13. In this paper, we consider the first 13 coefficients of MFCC and neglect the remaining coefficients. The average values of the above features for each emotion appear in Table II.

*B. Emotion classification*

We tested the speech emotion classification model using the Berlin Emo [4] database. At first, we trained the SVM model with 400 audio samples. Upon testing it with 135 new audio samples, the model resulted in an accuracy of 62.22%. In the second batch, we used 420 samples to train the SVM model. It resulted in an average accuracy of 68.25% upon testing with 115 audio samples. Finally, after training the SVM model with 440 audio samples, it resulted in an average accuracy of 78.94% upon testing with 95 audio samples[6].

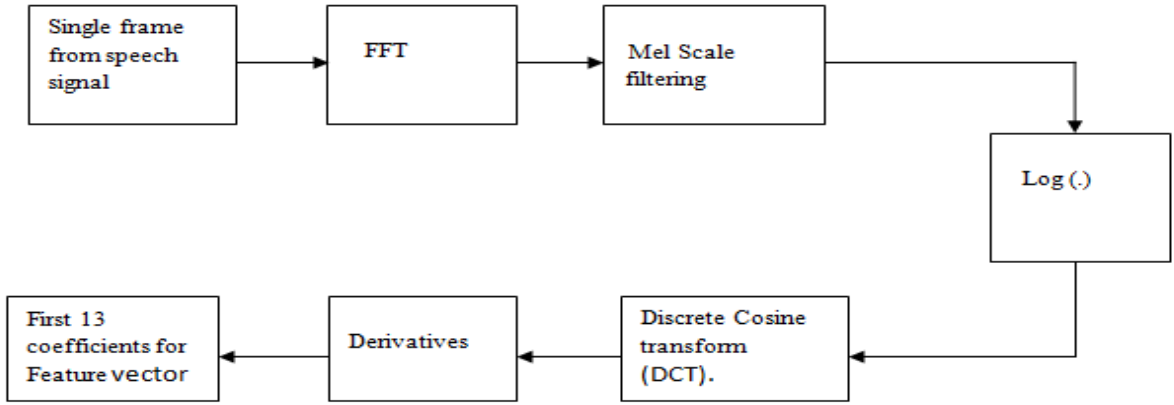

Fig. 13. MFCC block digram.

TABLE II. AVERAGE VALUES OF SPEECH FEATURES FOR DIFFERENT EMOTIONS .

| Emotion | Pitch | Energy | Entropy | Spectral Centroid | Spectral Spread | Spectral Entropy | Spectral RollOff | Spectral Flux |
|---|---|---|---|---|---|---|---|---|
| Happy | 285.7 | 1.70 | 3.12 | 0.48 | 0.411 | 0.95 | 0.997 | 0.015 |
| Neutral | 266.6 | 8.63 | 2.99 | 0.47 | 0.37 | 1.23 | 0.85 | 0.032 |
| Sad | 571.4 | 2.313 | 2.70 | 0.45 | 0.29 | 0.692 | 0.94 | 0.0027 |
| Angry | 400.0 | 8.64 | 3.10 | 0.55 | 0.352 | 2.01 | 0.56 | 0.09 |
| Disgust | 93.4 | 7.60 | 2.98 | 0.485 | 0.15 | 1.204 | 0.82 | 0.002 |
| Fear | 150.9 | 8.20 | 2.62 | 0.386 | 0.48 | 0.337 | 0.87 | 0.0674 |
| Surprise | 137.9 | 1.17 | 2.64 | 0.491 | 0.65 | 0.652 | 0.75 | 0.025 |

## IV. DECISION AT THE COMBINING STAGE

For the given input video clip, we extract the audio and video clips, and both (audio and video) signals are processed separately for emotion recognition. The video shows that each emotion frame counts (number of frames in which a particular emotion repeats in the video signal) are stored. The emotion with a maximum number of frame counts becomes the emotion of the video clip. Similarly, we classify the emotion when the audio signal is processed. At the combining stage, we make a decision. We consider the difference between the first two maximum frame counts. If the difference is more significant than a threshold value, then emotion from the video signal is taken as the overall emotion else emotion classified from an audio signal is taken as the overall emotion. The decision-taking appears in Fig. 14.[1].

## V. RESULTS

We employed the Enterface'05 [7] database for emotion recognition from audio and video signals. The database has a total of 44 speakers, and each speaker spoke approximately 30 sentences. Three hundred best video clips from the database are selected, and 225 (75%) video clips are taken for training, and 75(25%) video clips are taken for testing.

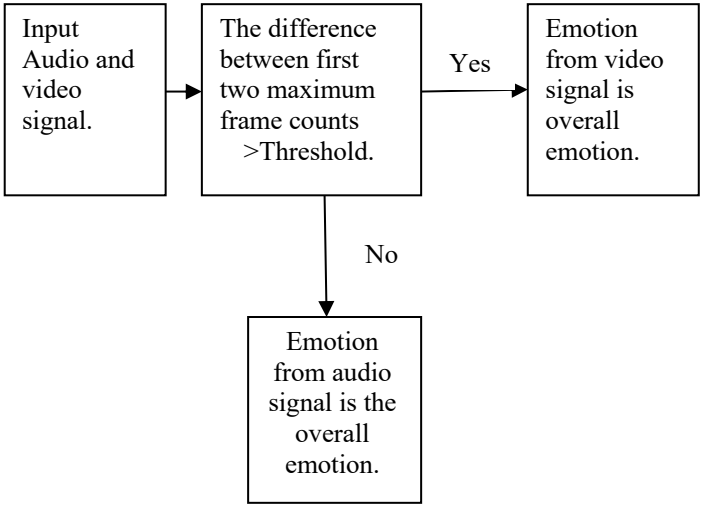

Fig. 14. Decision taking at the combining stage.

### A. Video accuracy

At first, we divided 225 videos into frames at a 24 FPS rate from the training set. Now image labeled database of 400 images is prepared manually from the frames of 225 video clips. We trained the image emotion classification model with this database. We divided the testing video into frames. From each frame, we predicted the emotion from the trained model. Now overall emotion carried throughout the video clip is the emotion that repeats in a maximum number of frames. We tested a total of 75 video clips, resulting in an accuracy of 74.66%. The confusion matrix appears in Table III.

### B. Audio accuracy

We extracted the audio clips from the training dataset. We trained all these 225 audio files to the speech emotion recognition model. We tested the remaining 75 audio files, and it resulted in an average accuracy of 69.33%. The confusion matrix appears in Table IV.

### C. Video and accuracy(overall)

At the combining stage, we set a threshold value for making a decision. By keeping the threshold value as nine, we achieved an average accuracy of 77.33%. We achieved different accuracies by varying the threshold value from 0 to 10. The plot of accuracy vs. threshold appears in Fig. 15. The confusion matrix appears in Table V.

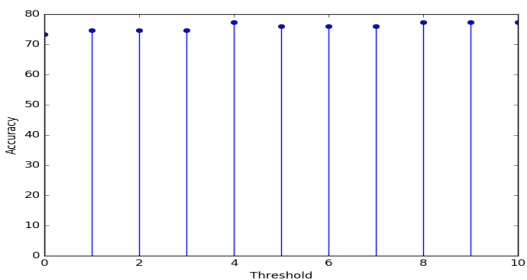

Fig. 15. Accuracy vs threshold values plot.

TABLE III. AVERAGE CONFUSION MATRIX FOR EMOTION RECOGNITION FROM VIDEO SIGNAL.

|    | An    | Ha    | Di    | Fe    | Sa    | Su    |
|----|-------|-------|-------|-------|-------|-------|
| An | 62.5  | 0     | 0     | 12.5  | 12.5  | 12.5  |
| Ha | 0     | 92.3  | 0     | 7.69  | 0     | 0     |
| Di | 18.18 | 27.27 | 54.54 | 0     | 0     | 0     |
| Fe | 0     | 0     | 10    | 60    | 10    | 10    |
| Sa | 5.55  | 0     | 0     | 5.55  | 83.33 | 5.55  |
| Su | 0     | 0     | 0     | 6.66  | 13.33 | 80.0  |

TABLE IV. AVERAGE CONFUSION MATRIX FRO EMOTION RECOGNITION FROM AUDIO SIGNAL.

|    | An   | Ha   | Di   | Fe    | Sa   | Su  |
|----|------|------|------|-------|------|-----|
| An | 75   | 25   | 0    | 0     | 0    | 0   |
| Ha | 7.69 | 46.5 | 7.69 | 23.07 | 0    | 15  |
| Di | 0    | 18.1 | 72.7 | 9.09  | 0    | 0   |
| Fe | 10   | 0    | 20   | 60    | 0    | 10  |
| Sa | 0    | 0    | 0    | 5.55  | 83.3 | 5.5 |
| Su | 0    | 0    | 20.0 | 6.66  | 6.66 | 66  |

TABLE V. AVERAGE CONFUSION MATRIX FRO EMOTION RECOGNITION FROM BOTH SIGNALS.

|    | An   | Ha    | Di    | Fe    | Sa    | Su   |
|----|------|-------|-------|-------|-------|------|
| An | 75   | 0     | 0     | 12.5  | 0     | 12.5 |
| Ha | 0    | 76.9  | 0     | 15.38 | 0     | 7.69 |
| Di | 9.09 | 18.18 | 72.72 | 0     | 0     | 0    |
| Fe | 0    | 0     | 0     | 70    | 10    | 20   |
| Sa | 5.55 | 0     | 0     | 0     | 88.88 | 5.55 |
| Su | 0    | 0     | 6.66  | 6.66  | 13.3  | 73   |

## VI. CONCLUSION

In this paper, we demonstrate a model for emotion recognition from audio and video signals that can be deployed in devices with nominal capabilities. The model relies on specific signal processing techniques employed to identify the emotion that are computer resource-friendly. Our complete model accuracy is 77.33%. Almost all digital voice assistants have sophisticated machine learning models incorporated in them. But to identify an emotion, for example, the data has to be transferred to the server where the emotion classification takes place. Our model (considering speech signal processing alone) can perform emotion classification without having to move data to servers. Similarly, the model can be used in dash-cameras in cars to alert a person if he/she is driving in an angry mood, for example. We conclude our research by demonstrating a model that can be deployed in customized devices for specific purposes.